\documentclass[%
 aip,
 amsmath,amssymb,
 reprint,%
]{revtex4-1}

\usepackage{graphicx}
\usepackage{dcolumn}
\usepackage{bm}

\usepackage[utf8]{inputenc}
\usepackage[T1]{fontenc}
\usepackage{mathptmx}

\begin{document}


\title{Spin Nematic Liquid of the S=1/2 Distorted Diamond Spin Chain in Magnetic Field}
\author{T\^oru Sakai}
\thanks{corresponding author}
\email[]{sakai@spring8.or.jp}
\altaffiliation{National Institutes for Quantum Science and Technology, SPring-8,
Hyogo 679-5148, Japan}
\author{Hiroki Nakano}
\author{Rito Furuchi}
\author{Kiyomi Okamoto}
\affiliation{ 
Graduate School of Science, University of Hyogo, Hyogo 678-1279, Japan}

\date{\today}

\begin{abstract}
The magnetization process of the $S=1/2$ distorted diamond spin chain 
with the anisotropic ferromagnetic interaction is investigated 
using the numerical diagonalization of finite-size clusters. 
It is found that the spin nematic and SDW Tomonaga-Luttinger liquids can appear 
for sufficiently large easy axis anisotropy. 
\end{abstract}


\maketitle

\def\vS{{\bf S}}  

\section{Introduction}

The spin nematic state is one of interesting phenomena in the field 
of the magnetism. 
It had been theoretically predicted to appear in the quantum spin systems 
with the biquadratic interaction or the spin frustration.\cite{andreev,chen,sudan,hikihara}
Recently another mechanism of the spin nematic liquid based on the 
anisotropy was proposed and it was predicted to occur in the 
spin ladder and the bond-alternating chain, 
using the numerical diagonalization analyses.\cite{sakai2010,sakai2020,sakai2021,sakai2022,sakai2022a,nakanishi}. 
In this paper, we propose the spin nematic liquid phase 
in the distorted diamond chain based on the same mechanism. 
\begin{figure}[tbh]
\includegraphics[width=0.7\linewidth]{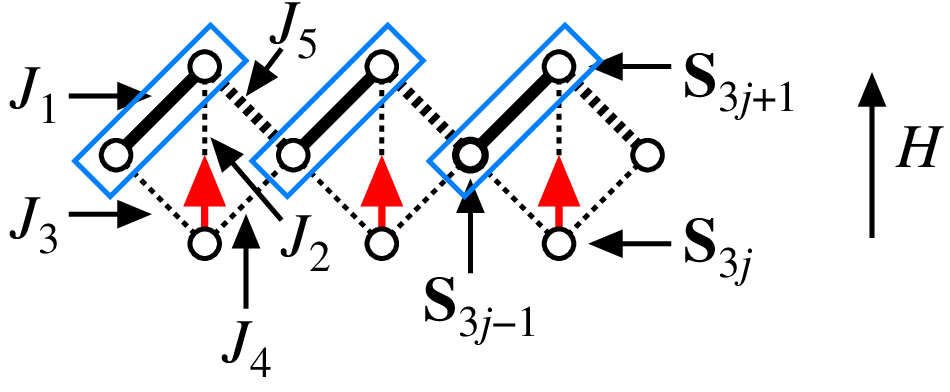}
\caption{The model of the $S=1/2$ distorted diamond spin chain. 
         Thick solid lines represent $J_1<0$, 
         thin dotted lines $J_2 = J_3 =J_4 >0$,
         and thick dotted lines $J_5>0$. 
         Blue rectangles and red arrows are for
         the explanation of the field-induced spin nematic Tomonaga-Luttinger state (see text).
}
\label{model}
\end{figure}

The distorted diamond chain is the theoretical model of 
the compound Cu$_3$(CO$_3$)$_2$(OH)$_2$, called azurite.\cite{kikuchi,okamoto1,okamoto2} 
In this compound all the exchange interaction are antiferromagnetic and 
it exhibits the 1/3 magnetization plateau due to the trimer nature. 
Recently other candidate materials for the distorted diamond spin chain were discovered. 
They are the compound  K$_3$Cu$_3$AlO$_2$(SO$_4$)$_4$, 
called alumoklyuchevskite\cite{fujihara,morita,fujihala2}
and related materials. 
This compound is also a frustrated system, but it includes 
some ferromagnetic interactions. 
Thus now we consider the distorted diamond chain including 
ferromagnetic interactions with the $XXZ$ anisotropy. 
The magnetization process of this model is investigated using the 
numerical diagonalization of finite-size clusters. 
As a result we will indicate that the field-induced 
spin nematic Tomonaga-Luttinger liquid phase can 
appear for some typical parameters. 
Its physical picture is shown by blue rectangles and red arrow in Fig. \ref{model}.
Namely, two spins in a blue rectangle are in 
$|\uparrow \uparrow \rangle$ or  $|\downarrow \downarrow \rangle$,
and behave as an effective $S=1/2$ spin having the doubled magnetic moment.
The spins in the lower column point to the magnetic field direction
as shown by red arrows.

\section{Model}

We investigate the model described by the Hamiltonian
\begin{eqnarray}
  {\cal H} &=& {\cal H}_0 + {\cal H}_{\rm Z} \\
  {\cal H}_0
&=& J_1 \sum _{j=1}^{N/3} \left[ S_{3j-1}^x S_{3j+1}^x + S_{3j-1}^y S_{3j+1}^y 
    + \lambda S_{3j-1}^z S_{3j+1}^z \right]\nonumber \\
   &&  +J_2 \sum_{j=1}^{N/3} \left[ \vS_{3j} \cdot \vS_{3j+1} \right] 
       +J_3 \sum_{j=1}^{N/3} \left[ \vS_{3j-1} \cdot \vS_{3j} \right] \nonumber \\
   &&    +J_4 \sum_{j=1}^{N/3} \left[ \vS_{3j} \cdot \vS_{3j+2} \right] 
         +J_5 \sum_{j=1}^{N/3} \left[ \vS_{3j-2} \cdot \vS_{3j-1} \right]  \\
         {\cal H}_{\rm Z}
   &=& -H \sum_{l=1}^{N} S_l^z 
   \label{eq:ham}
\end{eqnarray}
where $\vS_j$ is the spin-1/2 operator, $J_1$, $J_2$, $J_3$, $J_4$, $J_5$ are the 
coupling constants of the exchange interactions, respectively,
and $\lambda$ is the coupling anisotropy. 
The schematic picture of the model is shown in Fig. \ref{model}. 
In this paper we consider the case where $J_1$ is ferromagnetic 
and the Ising-like (easy-axis) anisotropy is introduced to this bond only 
($J_1<0$ and $\lambda > 1$), 
while $J_2$, $J_3$, $J_4$ and $J_5$ are isotropic antiferromagnetic bonds
($J_2, J_3, J_4, J_5 >0$ ). 
$N$ is the number of spins and $L$ is defined as the number 
of the unit cells, namely $N=3L$. 
For $L$-unit systems, 
the lowest energy of ${\cal H}_0$ in the subspace $\sum _j S_j^z=M$ is denoted as $E(L,M)$. 
The reduced magnetization $m$ is defined as $m=M/M_{\rm s}$, 
where $M_{\rm s}$ denotes the saturation of the magnetization, 
namely $M_{\rm s}=3L/2$ for this system. 
$E(L,M)$ is calculated by the Lanczos algorithm under the 
periodic boundary condition ($ {\bf S}_{N+l}={\bf S}_l$) 
for $L=4, 6$ and $8$. 

In order to consider the spin nematic liquid phase, 
we specify the parameters as follows: 
$J_1=-1.0$, $J_2=J_3=J_4=0.2$ and $J_5=0.5$. 
For these parameters, we will show that the field-induced 
spin nematic liquid phase appears for sufficiently large $\lambda$.

\section{Magnetization Curve}

\begin{figure}
\bigskip
\bigskip
\bigskip
\includegraphics[width=0.75\linewidth,angle=0]{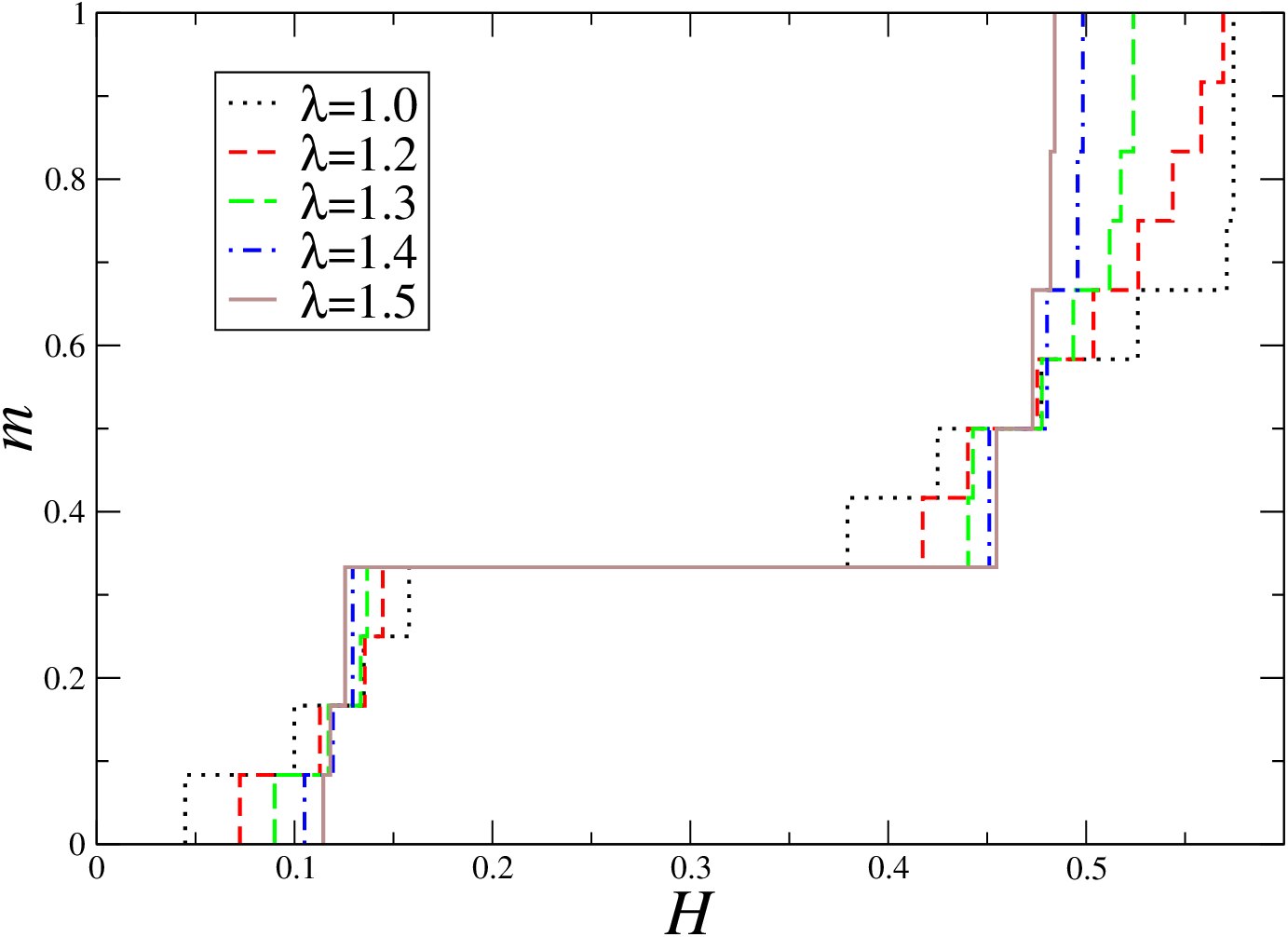}%
\caption{\label{mag}
Magnetization curves for $L=8$ calculated by the numerical diagonalization 
for $\lambda=$1.0, 1.2, 1.3, 1.4 and 1.5. 
 }
\end{figure}

The ground-state magnetization curves for $L=8$ are shown in 
Fig. \ref{mag} for $\lambda=1.0, 1.2, 1.3, 1.4$ and $1.5$. 
The 1/3 magnetization plateau clearly appears in all cases. 
The mechanism of the plateau is as follows: 
Since $J_2$, $J_3$, and $J_4$ are smaller than $J_1$ and $J_2$, 
the model behaves like the $J_1-J_5$ bond-alternating chain 
plus almost free spins at 3$j$ sites ($j=1, \cdots, L$). 
The 1/3 plateau is effectively the state where all the 3$j$ spins are up and 
the other spins form the ferromagnetic and antiferromagnetic alternating chain. 
Thus the magnetization process at $1/3 <m <1$ would correspond to 
the one of the bond-alternating chain. 
We look for the spin nematic liquid phase in this magnetization region. 

When $\lambda \sim 1$,
the ground-state of the $J_1-J_5$ chain will be the singlet dimer state
which is smoothly connected to the Haldane state of the spin-1 chain.\cite{hida}
In this case, the step of the magnetization curve is $\delta M=1$.
On the other hand, the Ising-like anisotropy $\lambda (> 1)$ stabilizes the states 
$|\uparrow \uparrow \rangle$ and $|\downarrow \downarrow \rangle $ 
at the $J_1$ bond (see Fig. \ref{model}). 
As a result each step of the magnetization curve tends to be $\delta M=2$. 
This two magnon bound state is one of the characters of the spin nematic 
liquid phase. 
In fact, some steps with $\delta M=2$ appear in the magnetization curves 
for $\lambda \ge 1.3$ in Fig. \ref{mag}. 
Thus we investigate the possibility of the spin nematic liquid 
around the magnetization region where the two magnon bound state 
is realized.

\section{1/3 Magnetization Plateau}

At the 1/3 magnetization plateau the N\'eel order of the $J_1-J_5$ chain
is expected to occur 
for sufficiently large $\lambda$. 
The phase transition from the singlet dimer state to the N\'eel ordered state is
of the same universality class of that from the Haldane state to the N\'eel state
of the spin-1 chain.
Using the phenomenological renormalization group analysis,\cite{nightingale} the critical point 
$\lambda_{\rm c}$ of the quantum phase transition can be estimated. 
The size-dependent fixed point is determined by the form
\begin{eqnarray}
(L+2)\Delta_{\pi}(L+2, \lambda)=L\Delta_{\pi}(L,\lambda),
\label{prg}
\end{eqnarray}
where $\Delta _{\pi}(L,\lambda)$ is the gap of the
excitation with $k=\pi$ at $m=1/3$. 
The scaled gap $L\Delta _{\pi}(L,\lambda)$ is plotted versus $\lambda$ 
for $L=$2, 4, 6 and 8 in Fig. \ref{prg}(a). 
The size-dependent fixed point $\lambda_{\rm c}(L+1)$ determined as the cross point 
of $L\Delta _{\pi}(L,\lambda)$ and $(L+2)\Delta_{\pi}(L+2,\lambda)$ 
is plotted versus $1/(L+1)^2$ in Fig. \ref{prg}(b). 
Assuming the size correction is proportional to $1/(L+1)^2$, 
the critical point in the infinite $L$ limit is estimated 
as $\lambda_{\rm c} = 1.342 \pm 0.001$.

\begin{figure}[tbh]
\bigskip
\bigskip
\includegraphics[width=0.8\linewidth]{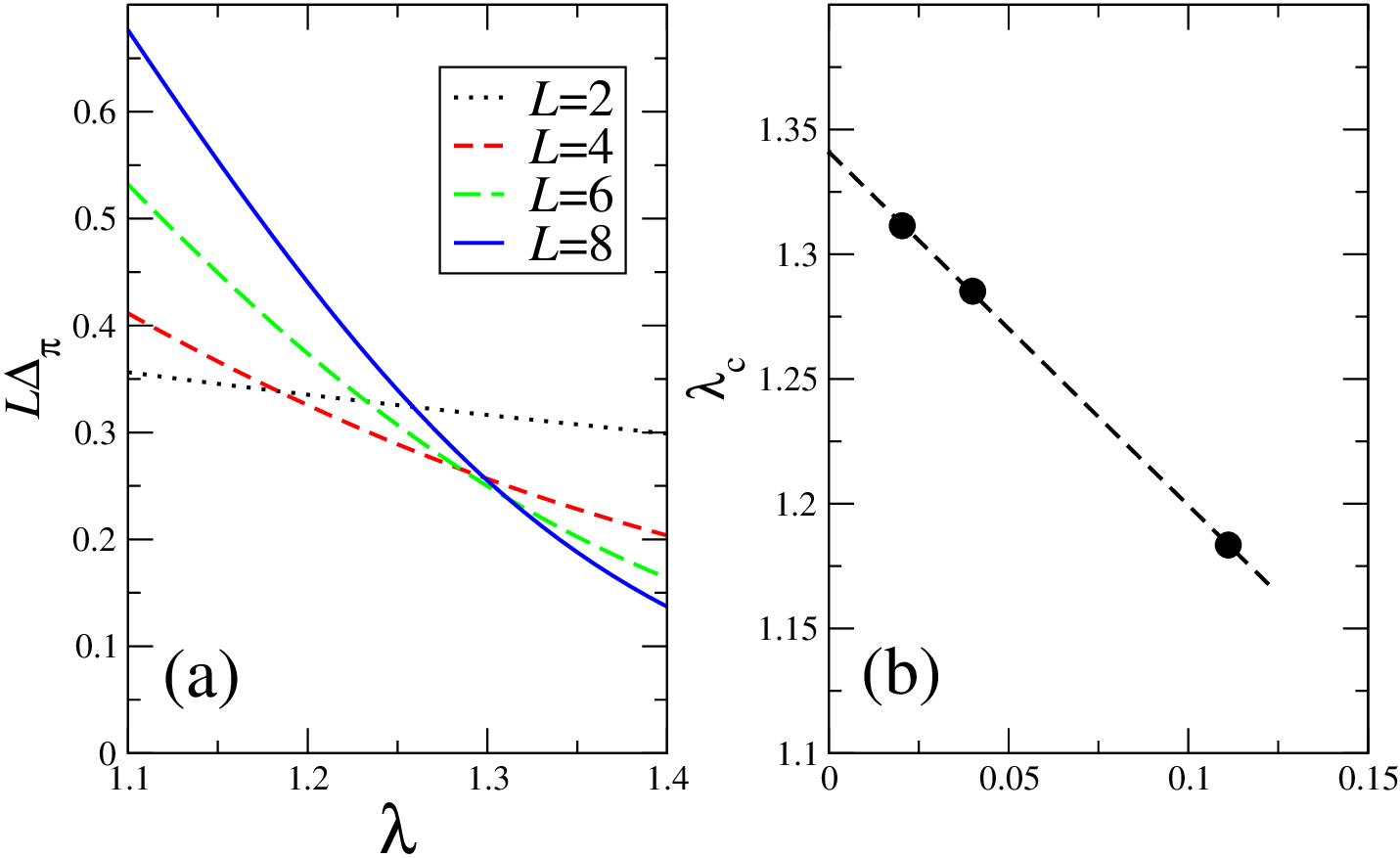}
\caption{
(a) Scaled gaps $L\Delta _{\pi}(L, \lambda)$ are plotted versus $\lambda$ for 
$L=$2, 4, 6, and 8. 
(b) Extrapolation of the size-dependent fixed points $\lambda _c (L+1)$ 
to the infinite $L$ limit, assuming the size correction is proportional 
to $1/(L+1)^2$. 
}
\label{prg}
\end{figure}

\section{Two Tomonaga-Luttinger liquids}

The gapless region in the magnetization process of the quantum spin chain 
is generally in the Tomonaga-Luttinger liquid\cite{haldane} with $\delta M=1$. 
We call it the conventional Tomonaga-Luttinger liquid (CTLL). 
The present model is in the CTLL phase around $\lambda=1$. 
As shown in the section III, the two magnon bound state with $\delta M=2$ 
can be realized due to the sufficiently large $\lambda$. 
This phase is called the two-magnon TLL phase. 
The two-magnon TLL phase by similar mechanism was found
in several models.\cite{sakai2010,sakai2020,sakai2021,sakai2022,sakai2022a,nakanishi}. 

\begin{figure}
\includegraphics[width=0.85\linewidth,angle=0]{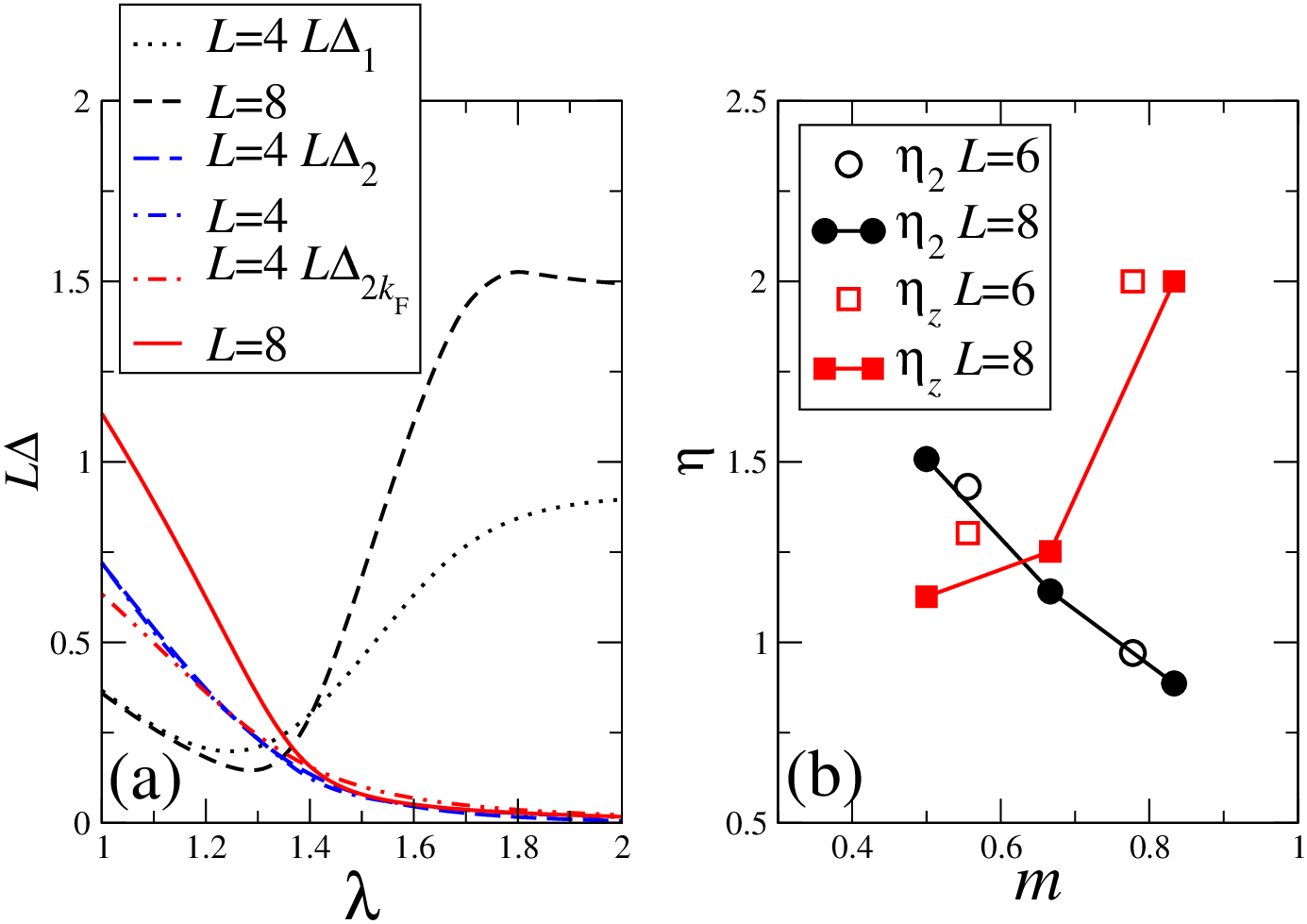}%
\caption{\label{sg}
(a) Scaled gaps $L\Delta _1$, $L\Delta _2$ and $L\Delta _{2k_F}$ are plotted versus 
$\lambda$ for $L=$4 and 8 for $m=2/3$. 
(b) Exponents $\eta_2$ and $\eta_z$ estimated for $L=6$ and 8 are plotted versus $m$ for $\lambda =1.5$. 
 }
\end{figure}

The behaviors of several excitation gaps are different 
between these two TLL phases. 
The single- and two-magnon excitation gaps are denoted as $\Delta_1$ and 
$\Delta_2$. In addition the $2k_F(=3/2(1-m)\pi)$ excitation gap of the 
two magnon bound state is denoted as $\Delta _{2k_F}$. 
The scaled gaps $L\Delta _1$, $L\Delta _2$ and $L\Delta _{2k_F}$ are plotted versus 
$\lambda$ for $L=$4 and 8 for $m=2/3$. 
The small (large) $\lambda$ region is in the CTLL (two-magnon TLL) phases. 
Fig. \ref{sg} indicates that $\Delta_1$ ($\Delta _{2k_F}$) is gapless (gapped) 
in the CTLL phase, while $\Delta_1$ ($\Delta _{2k_F}$) is gapped (gapless) 
in the two-magnon TLL one. 
In contrast, $\Delta_2$ is always gapless. 
Thus the cross point of $\Delta_1$ and $\Delta_{2k_F}$ 
and that of $\Delta_1$ and $\Delta_2$ are good probes for the estimation of
the phase boundary between the CTLL and the two-magnon TLL phases. 
Here we adopt the former from the viewpoint of the finite-size effect.\cite{sakai2022a,nakanishi}

In the two-magnon TLL phase, 
the quasi-long-range SDW and nematic orders are expected to be realized. 
They are characterized by the power law decays of the following spin 
correlation functions
\begin{eqnarray}
\langle S_{1}^zS_{3r+1}^z \rangle -\langle S_1^z \rangle^2 & \sim & \cos(2k_{\rm F}r)r^{-\eta_z} 
\label{sdw}\\
\langle S_{1}^+S_{2}^+ S_{3r+1}^-S_{3r+2}^- \rangle & \sim & r^{-\eta_2},
\label{nematic}
\end{eqnarray}
where the first equation corresponds to the SDW spin correlation parallel to the external field 
and the second one corresponds to the nematic spin correlation perpendicular to the external field. 
The smaller exponent between $\eta_z$ and $\eta_2$ determines the dominant spin correlation. 
According to the conformal field theory these exponents can be estimated by the forms\cite{cardy}
\begin{eqnarray}
\eta_2&=&{{E(L,M+2)+E(L,M-2)-2E(L,M)}\over{E_{k_1}(L,M)-E(L,M)}}, \\
\eta_z&=&2{{E_{2k_F}(L,M)-E(L,M)}\over{E_{k_1}(L,M)-E(L,M)}},
\label{exponent}
\end{eqnarray}
for each magnetization $M$, where $k_1$ is defined as $k_1=L/2\pi$. 
The exponents $\eta_2$ and $\eta_z$ estimated for $L=6$ and 8 are plotted versus 
$m$ for $\lambda =1.5$ in Fig. \ref{sg}(b). 
It suggests that the SDW correlation is dominant for small $m$, 
while the nematic one for large $m$. 
Neglecting the finite size correction, the cross point of $L=8$ is used as the 
crossover point between the nematic correlation dominant TLL (NTLL) and the 
SDW correlation dominant TLL (SDWTLL) phases.

\section{Phase Diagram}
\begin{figure}
\includegraphics[width=0.75\linewidth,angle=0]{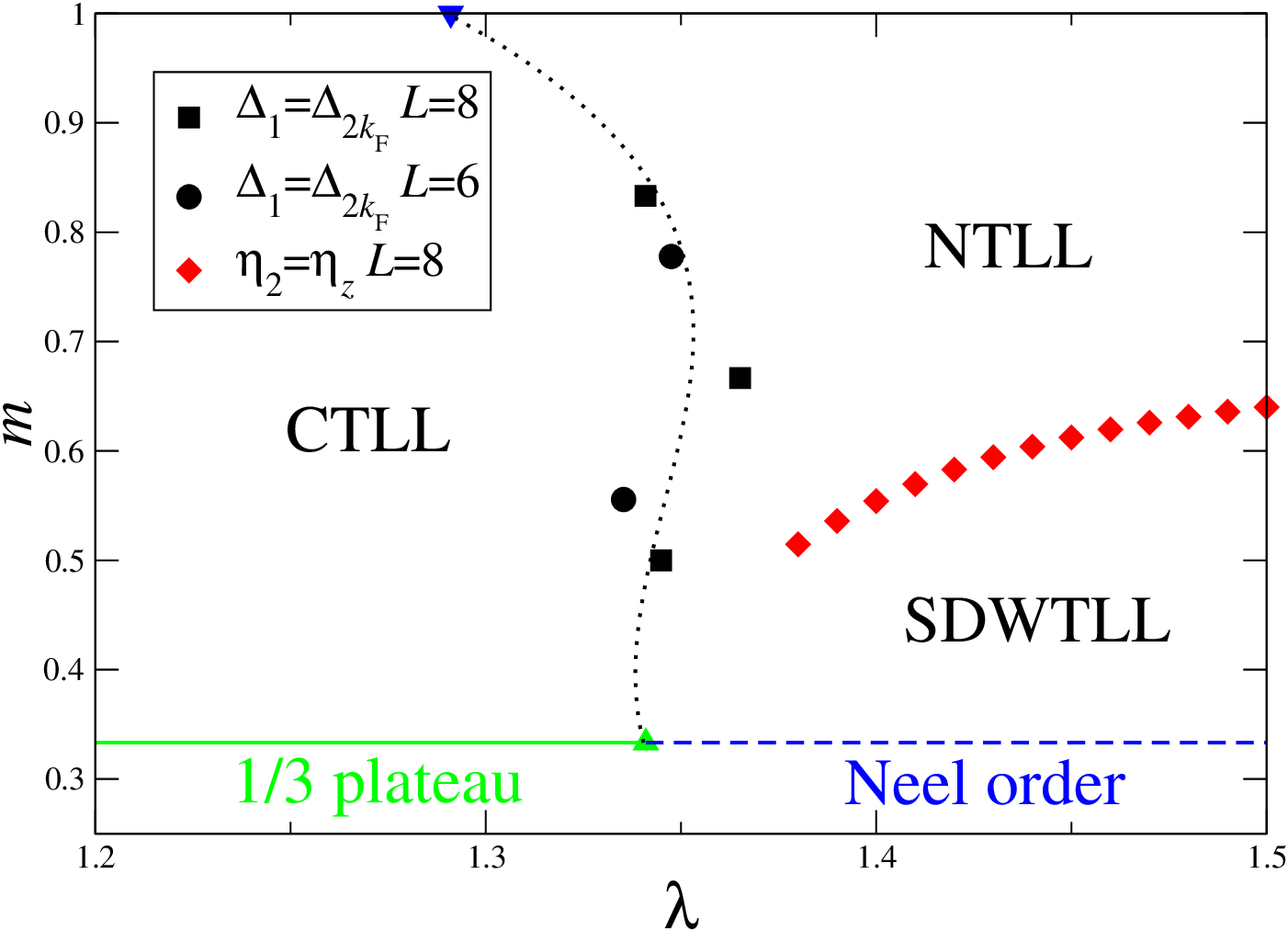}%
\caption{\label{phase}
Phase diagram with respect to the anisotropy $\lambda$ and the magnetization $m$.} 
\end{figure}

Finally we obtain the phase diagram with respect to the anisotropy $\lambda$ and the 
magnetization $m$ in Fig. \ref{phase}. 
The phase boundary between the CTLL and the two-magnon TLL phases is estimated as 
the cross point between $\Delta_1$ and $\Delta_{2k_F}$ for $L=6$ and 8 (solid circles 
and squares, respectively). 
The crossover line is determined by $\eta_2=\eta_z$ (red diamonds). 
At the 1/3 magnetization plateau the boundary of the N\'eel ordered phase 
is determined by the phenomenological renormalization (up triangle). 
This boundary is expected to be connected to the one between the CTLL and two-magnon 
TLL phases. 
The boundary at $m=1$ can be easily determined by $\Delta_1=\Delta_2$. 
The magnetization process at $0<m<1/3$ is also expected to be in the TLL phase. 
However, whether some multi-magnon TLL phases possibly appear or not 
is still an unsolved problem.

\section{Summary}

The $S=1/2$ distorted diamond spin chain with the anisotropic ferromagnetic interaction is investigated 
using the numerical diagonalization. 
As a result it is found that as the nematic correlation dominant TLL phase 
can appear for a sufficiently large anisotropy, as well as the SDW correlation dominant TLL phase. 
The phase diagram with respect to the anisotropy and the magnetization 
is also presented.

\begin{acknowledgments}
This work was partly supported by JSPS KAKENHI, 
Grant Numbers JP16K05419, JP20K03866, JP16H01080 (J-Physics), 
JP18H04330 (J-Physics) and JP20H05274.
A part of the computations was performed using
facilities of the Supercomputer Center,
Institute for Solid State Physics, University of Tokyo,
and the Computer Room, Yukawa Institute for Theoretical Physics,
Kyoto University. 
We also used the computational resources of the supercomputer 
Fugaku provided by the RIKEN through the HPCI System 
Research projects (Project ID: hp200173, hp210068, hp210127, 
hp210201, and hp220043). 
\end{acknowledgments}

\section*{Data Availability}
The data that support the findings of this study are available from the corresponding author upon reasonable request.

\nocite{*}
\bibliography{aipsamp}

\end{document}